\documentclass[a4paper,11pt]{article}
\pdfoutput=1 
\usepackage{jinstpub}
\usepackage{enumerate}

\title{\bf Effect of electrical properties of glass electrodes on the performance of RPC detectors for the INO-ICAL experiment}

\author[a,c,1]{K. Raveendrababu, \note{Corresponding author.}}
\author[a]{P. K. Behera,}
\author[b]{B. Satyanarayana}

\affiliation[a]{Physics Department, Indian Institute of Technology Madras, Chennai 600036, India}
\affiliation[b]{Department of High Energy Physics, Tata Institute of Fundamental Research, Mumbai 400005, India}
\affiliation[c]{Physical Science Division, Homi Bhabha National Institute, Anushaktinagar, Mumbai 400094, India}

\emailAdd{ravi2ramana@gmail.com}

\abstract{
The India-based Neutrino Observatory (INO) collaboration has chosen glass Resistive Plate Chambers (RPCs) as the active detector elements for
the Iron Calorimeter (ICAL) experiment. In the present work, we study the electrical properties such as bulk resistivity and relative 
permittivity of the glasses from two different manufacturers and compared the performances of RPCs built using these glasses. We 
conclude that the glass electrodes with larger bulk resistivy and permittivity are better suited for manufacturing RPCs 
for the ICAL experiment, as these detectors could be operated at lower bias currents and voltages, and produce better time resolutions 
compared to those built with glass electrodes of smaller bulk resistivity and permittivity.   
}


\begin{document}
\maketitle
\flushbottom

\section{Introduction}
\label{Sec:Introduction}
\vskip 0.2cm
{Resistive Plate Chambers (RPCs) are gaseous detectors, which work on the ionization principle. They are simple to construct, offer 
two-dimensional readout, provide good efficiency (>90\%) and time resolution ($\sim$1 ns). Therefore, they are used as the trigger and/or 
timing detectors in many high-energy physics experiments~\cite{G_Bruno, Satyanarayana}.} 

\subsection{The INO-ICAL experiment} 
\label{Sec:The INO-ICAL experiment}
{The India-based Neutrino Observatory (INO) collaboration has proposed to build a 50 kiloton magnetized Iron Calorimeter (ICAL). The main aims of 
this experiment are to precisely measure the atmospheric neutrino oscillation parameters and to determine the ordering of neutrino masses. 
The ICAL detector will comprise of three modules of 16 $\times$ 16 $\times$ 14.5 m$^3$. Each module will consist of a stack of 151 horizontal 
layers of 5.6 cm thick iron plates interleaved with 4 cm gaps to house the active detector layers. The collaboration has chosen glass RPCs of 
2 $\times$ 2 m$^2$ in size as the active detector elements. The ICAL detector is going to use $\sim$28,800 such RPCs. The experiment is 
expected to operate for more than 10 years in order to record statistically significant number of neutrino interactions to study neutrino 
oscillations~\cite{Shakeel_Ahmed_et_al}. Therefore, long-term operation and performance of RPCs over the duration of the experiment are of 
prime concern. The ICAL requires a good detector time resolution ($\sim$1 ns) to discriminate the direction of particles traversing through 
its volume. The layout of the proposed modular form of the ICAL detector and its construction sequence are shown in figure~\ref{fig:1}{a} and 
\ref{fig:1}{b}, respectively~\cite{Satyanarayana, A_Behere_et_al}.

\begin{figure}[ht]
        \centering
         \includegraphics[width=0.7\textwidth]{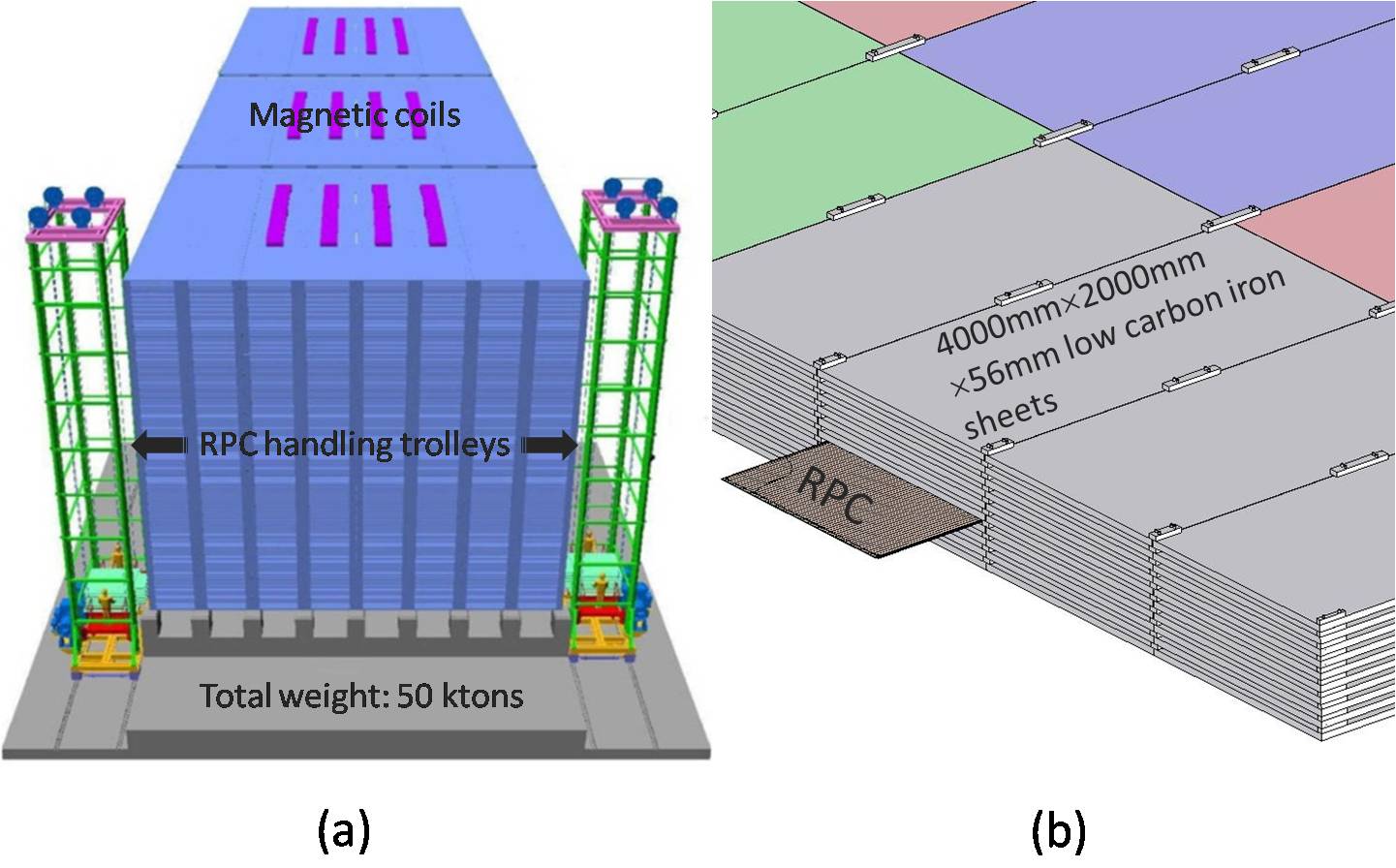}
          \caption {(a) The layout of the proposed INO-ICAL detector and (b) the construction sequence of the ICAL detector.}
\label{fig:1}
\end{figure}
}

\subsection{Resistive plate chamber} 
\label{Sec:Resistive plate chamber}
\vskip 0.2cm
{The RPC electrodes are made out of a high resistive material (typically, bakelite or float glass), with a  bulk resistivity of 
$10^{10} - 10^{12} \ \Omega$.cm. The outer surfaces of the electrodes are coated with a semi-resistive paint, which facilitated applying
high voltage across the electrodes~\cite{Santonico_Cardarelli, Archana_Sharma}. The electrodes are separated by a few millimeters using 
poly-carbonate spacers to maintain a uniform gap between the two electrodes. The electrodes are applied with high voltages $\sim$10 kV in order 
to create a uniform electric field in the gap. Conductive readout-strips are orthogonally mounted on the external surfaces of the gap. The 
electrodes and the readout-strips are separated using a layer of mylar insulator. The basic construction of RPC is shown in figure~\ref{fig:2}.

An optimized gas mixture of Ar/C$_{2}$H$_{2}$F$_{4}$/iso-C$_{4}$H$_{10}$ = 30/62/8~\citep{Satyanarayana, Yamaga_et_al} or C$_{2}$H$_{2}$F$_{4}$/iso-C$_{4}$H$_{10}$/SF$_{6}$ = 95.2/4.5/0.3~\cite{Salim_et_al} is flown through the RPC gap, depending on either streamer or avalanche mode of operation, respectively. When a charged particle passes through the gas gap, it produces the free charge carriers. These free charge 
carriers multiply into an avalanche under the influence of strong electric field in the gap, propagate toward the electrodes and induce charge 
on the external readout-strips. Since the readout-strips are orthogonally placed on either side of the gas gap, we get the (x, y) coordinates 
of the particle using the same active detector volume.

\begin{figure}[ht]
        \centering
         \includegraphics[width=1\textwidth]{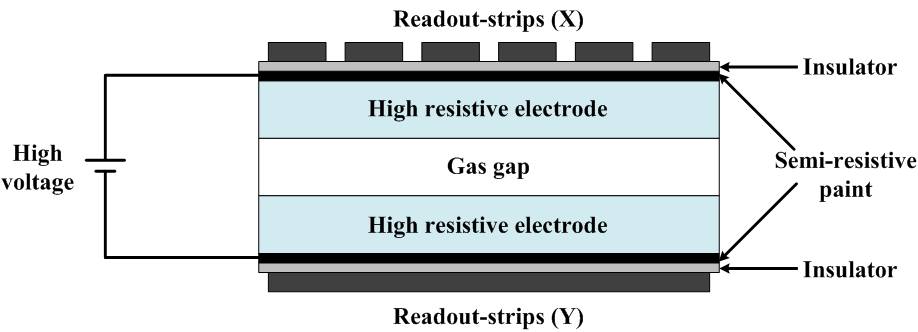}
          \caption {Basic construction schematic of the Resistive Plate Chamber.}
\label{fig:2}
\end{figure}

For the INO-ICAL experiment, the collaboration has proposed to use glasses from Asahi and/or Saint-Gobain for producing RPC gas gaps. The 
float glass is a dielectric material, hence operation and performance of the glass RPCs are influenced by bulk resistivity and relative 
permittivity of the electrodes. Therefore, here, we performed systematic studies on these electrical properties of the glass samples from 
Asahi and Saint-Gobain manufacturers and compared the performances of RPCs built using these glasses.
}

\section{Electrical property measurements on the glasses}

In an attempt to remove the surface contaminants, the glass samples were first treated as follows:

\begin{enumerate}[i.]
 \item thoroughly cleaned with the solution of labolene soap and deionized water,
 \item rinsed in deionized water,
 \item rinsed in iso-propyl alcohol, and
 \item finally dried in a filtered airstream.
\end{enumerate}

During and after this treatment the glass samples were handled on edges with latex free and powder free surgical gloves. The following 
measurements were performed on theses samples in a clean environment.

\subsection{Bulk resistivity}

The bulk resistivity of the glass samples was measured using Agilent Technologies B1500A Semiconductor Device Analyzer equipped with B1511B
Medium power source/monitor unit (MPSMU). The instrument provides accurate and precise measurement ranges of 10 fA - 0.1 A and 0.5 $\mu$V - 
100 V. Using the equipment, the voltage-current (V-I) characteristics of the glasses were measured on 2 $\times$ 2 cm$^{2}$ samples of 3 mm
thick. The volume resistance of the glasses was obtained from the slope of that V-I plots. Then, the volume or bulk resistivities $\rho$  
were obtained using the below relation:

\begin{equation}
 R = \frac{\rho . \ t}{A} \implies \rho = \frac{R \ . \ A}{t}, 
\label{eq:2.1}
\end{equation}

\noindent
where, $R$, $A$ and $t$ are the volume resistance, area and thickness of the glass sample, respectively~\cite{Dai_et_al}.

The measurements were taken on three samples of each manufacturer and the results were averaged out. The bulk resistivity measurements on the 
glass samples are summarized in table~\ref{tab:1}. The calculated standard deviations on the average bulk resistivities of Asahi 
and Saint-Gobain glasses ($\sigma_{\text{\bf A}}$ and $\sigma_{\text{\bf S}}$) are also given in the table. Asahi glass showed larger bulk 
resistivity compared to Saint-Gobain glass.

\begin{table}[ht]
\caption{Bulk resistivities of Asahi and Saint-Gobain glasses in $\Omega$.cm. $\sigma_{\text{\bf A}}$ and $\sigma_{\text{\bf S}}$ are the calculated standard deviations on the average bulk resistivities of Asahi and Saint-Gobain glasses, respectively.}
\begin{center}
\vskip -0.5cm
\resizebox{\textwidth}{!}{
\begin{tabular}{|c|c|c|c|c|}
\hline
{\bf Sample number} & {\bf Bulk resistivity of } & {$\mathbf{\sigma_{A}}$} & {\bf Bulk resistivity of} & {$\mathbf{\sigma_{S}}$} \\ 
                    & {\bf Asahi glass}         & & {\bf Saint-Gobain glass}  & \\
                    & {($\times$ 10$^{12}$ $\Omega$.cm)} & {($\times$ 10$^{12}$ $\Omega$.cm)} & {($\times$ 10$^{12}$ $\Omega$.cm)} & {($\times$ 10$^{12}$ $\Omega$.cm)}  \\
\hline
1 & 4.67 & & 3.87 & \\

2 & 4.72 & & 3.60 & \\

3 & 4.80 & & 3.47 & \\
\hline
Average & 4.73 & 0.05 & 3.65 & 0.17 \\
\hline
\end{tabular}
}
\label{tab:1} 
\end{center} 
\end{table}

\subsection{Relative permittivity}
\label{Sec:Relative permittivity}
{
The permittivity ($\epsilon$) characterizes the tendency of charge distortion or the electric polarization of a dielectric 
material (insulator) in the presence of an electric field. The larger permittivity implies the higher tendency of electric polarization.
The permittivity of a dielectric material is expressed as~\cite{Griffiths}:

\begin{equation}
\epsilon = \epsilon_{r} \epsilon_{0}, 
\label{eq:2.2}
\end{equation}

\noindent
where, $\epsilon_{0}$ is the permittivity of the vacuum and $\epsilon_{r}$ is the relative permittivity of the dielectric.   

The relative permittivities of the glass samples were measured using 'Novocontrol Broadband Dielectric/Impedance Spectrometer' at 
$20 \ ^{\circ}$C. The results are shown in figure~\ref{fig:3}. Asahi glass showed larger relative permittivity compared to 
Saint-Gobain glass. Hence, investigations were made to understand the reason for the difference in relative permittivity between the two 
glasses. Elemental compositions of the glass samples were measured using Energy Dispersive X-ray Spectroscopy detector equipped to FEI Quanta 200 
Scanning Electron Microscope. These results in fractional atomic percentages are summarized in table~\ref{tab:2}.

\begin{figure}[ht]
        \centering
         \includegraphics[width=0.55 \textwidth]{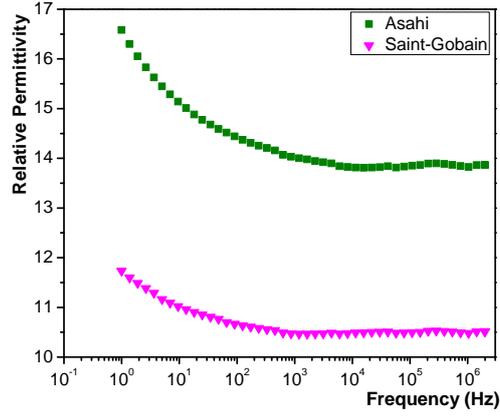}
         \vskip -0.5cm
          \caption{Relative permittivities of Asahi and Saint-Gobain glasses using Novocontrol Broadband 
                   Dielectric Impedance Spectrometer.}
\label{fig:3}
\end{figure}

Normally, the float glass is made out of the chemical composition SiO$_{2}$, Na$_{2}$O, MgO, Al$_{2}$O$_{3}$ and CaO. The Si$^{4+}$ 
ions are called as network-formers and the corresponding bonding oxygen ions as $bridging \ oxygens$. The formation of SiO$_{2}$ 
(Si -- O -- Si) network in glass is illustrated in figure~\ref{fig:4}{a}. The Na$^{+}$, Mg$^{2+}$, Al$^{3+}$ and Ca$^{2+}$ ions 
are called as network-modifiers and the corresponding bonding oxygen ions as $nonbridging \ oxygens$. The modification of SiO$_{2}$ network 
through Na$^{+}$ (alkali ion) network modifiers is illustrated in figure~\ref{fig:4}{b}. In an electric field, the more content 
of network modifiers leads to more easily polarizable $nonbridging \ oxygens$ and therefore increase of the relative permittivity of material
\cite{Hsieh_Jain_Kamitsos, Horst_Scholze}. The tested Asahi glass has larger amount of sodium (Na) component ($\sim$ 2\%) compared to Saint-Gobain glass as 
seen in table~\ref{tab:2}. This could be the reason for Asahi glass showing larger relative permittivity compared to Saint-Gobain glass.

\begin{table}[ht]
\caption{Summary of the various elemental composition of glass samples in fractional percentages of atoms (\%).}
\begin{center}
\vskip -0.8cm
\Large
\resizebox{\textwidth}{!}{
\begin{tabular}{|l|c|c|c|c|c|c|c|}
\hline
{\bf Element} & \multicolumn{3}{c|}{\bf Asahi (\%)} & \multicolumn{3}{c|}{\bf Saint-Gobain (\%)} & {\bf A1--A2 (\%)}\\ 
\cline{2-7}
&\bf Sample-1 &\bf Sample-2 &\bf Average (A1) &\bf Sample-1 &\bf Sample-2 &\bf Average (A2) & \multicolumn{1}{c|}{} \\
\hline
Oxygen (O) & 70.93 & 68.65 & 69.79 & ~71.03 & 71.38 & 71.21 & ~-- 1.42 \\

Silicon (Si) & 15.43 & 16.69 & 16.06 & 16.7 & 16.86 & 16.78 & ~-- 0.72 \\

Sodium (Na) & 10.46 & 10.71 & 10.59 & ~8.7 & ~8.29 & 8.5 & + 2.09 \\

Magnesium (Mg) & ~2.21 & ~2.51 & ~2.36 & ~~2.17 & ~2.31 & ~2.24 & + 0.12 \\

Calcium (Ca) & ~0.48 & ~0.82 & ~0.65 & ~~0.85 & 0.6 & ~0.73 & ~-- 0.08 \\

Aluminium (Al) & ~0.48 & ~0.56 & ~0.52 & ~~0.46 & ~0.52 & ~0.49 & + 0.03 \\

Iron (Fe) & ~0.02 & ~0.05 & ~0.04 & ~~0.08 & ~0.04 & ~0.06 & ~-- 0.02 \\
\hline
\end{tabular}
}
\label{tab:2} 
\end{center} 
\end{table} 

\begin{figure}[ht]
        \centering
         \includegraphics[width=0.55 \textwidth]{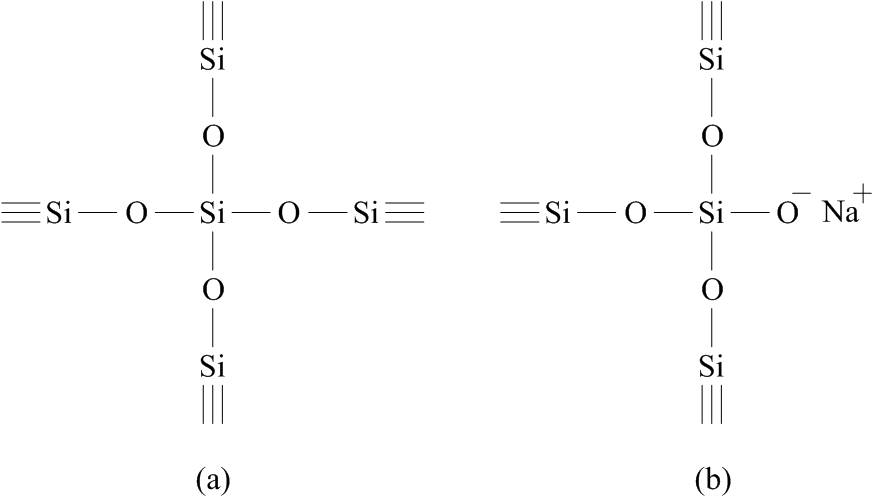}
          \caption{Schematic diagrams of (a) the Si -- O -- Si network formation, and (b) the Si -- O -- Si network modification through 
                   Na$^{+}$ ions in float glass.}
\label{fig:4}
\end{figure}

The relaxation time of the RPC electrode is $\tau$ = RC = $\rho \epsilon_0 \epsilon_r$. Therefore as per the relative permittivity measurements 
in figure~\ref{fig:3}, we expect that under identical conditions Saint-Gobain RPCs offer better counting rate capability compared to Asahi
RPCs. Even though, it is not of consequence in the case of INO-ICAL experiment, where the expected counting rate is much lower.

From~\cite{Bilki_et_al}, it is understood that the induced charge on the strip is inversely proportional to the relative permittivity 
($\epsilon_{r}$) of the electrode for the same amount of charge produced in the avalanche. The analog pulses from Asahi RPC at 11 kV 
and Saint-Gobain RPC at 12.2 kV were recorded on Agilent DSO-X 3104A. The selected operating voltages were 200 V upstream of the knee of 
efficiency plateau (figure~\ref{fig:9}) of each RPC, respectively. These pulses were integrated and measured the collected charge per event 
of RPCs. The results are shown in figure~\ref{fig:5}. The obtained mean values from the charge distributions of Asahi and Saint-Gobain RPCs
are 1.53 and 1.86, respectively. The ratio of these means, i.e., 1.86/1.53 is close to the inverse ratio of the relative permittivities of the 
corresponding electrodes, i.e., 14/10.5 (figure~\ref{fig:3}).    

\begin{figure}[ht]
        \centering
         \includegraphics[width=0.9\textwidth]{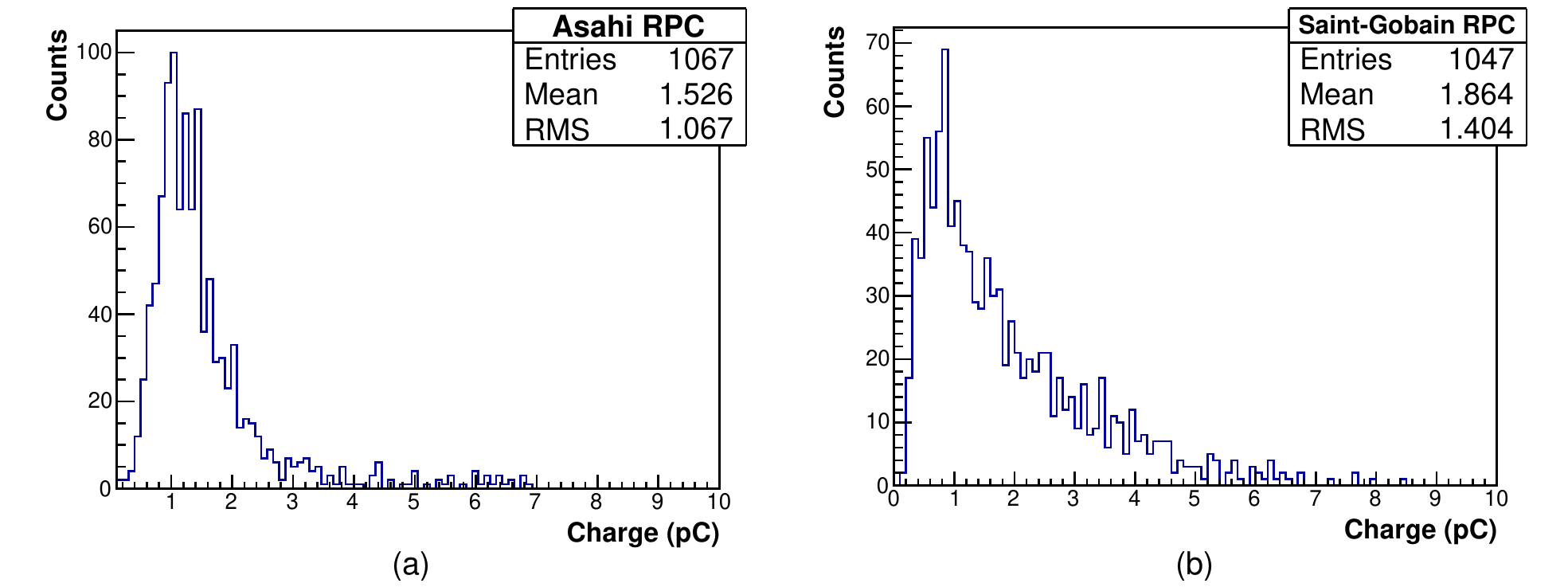}
         \vskip -0.3cm
          \caption {Charge per event of: (a) Asahi RPC at 11 kV and (b) Saint-Gobain RPC at 12.2 kV.}
\label{fig:5}
\end{figure}
}

\section{RPC performance studies}
\label{Sec:RPC performance studies}
{
RPCs of 30 $\times$ 30 cm$^2$ size were built using 3 mm thick float glass plates from both the manufacturers. The outer surfaces of 
the glass plates were coated with specially developed conductive graphite paint for the ICAL RPCs~\cite{Bhuyan_et_al}. The surface resistance 
of the coated film was maintained to be uniform at $\sim$1 M$\Omega$/ \hspace{-5mm}$\qed$. The parallel electrodes were separated by 0.2 cm using 
polycarbonate spacers and gas nozzles, whose bulk resistivity is $\sim$10$^{13} \ \Omega$.cm~\cite{Satyanarayana}. Finally, the outer spacers were 
glued to the electrodes using 3M Scotch-Weld DP190 Epoxy Adhesive to isolate the detector volume from the ambient environmental conditions.  
Readout-strips of 2.8 cm wide, with 0.2 cm gap between the consecutive strips, were orthogonally mounted on the external surfaces of the RPCs. A layer of mylar sheet was inserted between the electrode and the readout-strip. The gaps made out of Asahi glass and Saint-Gobain glass were named as A-RPCs and S-RPCs, respectively. A gas mixture of C$_{2}$H$_{2}$F$_{4}$/iso-C$_{4}$H$_{10}$/SF$_{6}$ = 95/4.5/0.5 was flown through the RPCs with a total flow rate of 10 SCCM and operated in the avalanche mode. All the RPCs were operated under identical environmental conditions.

\subsection{Experimental setup}
\label{Sec:Experimental setup}
{
A cosmic ray muon telescope was set up with three plastic scintillation counters to get a 3-fold coincidence. The dimensions of scintillation 
counters in length $\times$ width $\times$ thickness are 30 $\times$ 2 $\times$ 1 cm$^3$ (top), 30 $\times$ 3 $\times$ 1 cm$^3$ (middle), and 
30 $\times$ 5 $\times$ 1 cm$^3$ (bottom). The RPCs were stacked between top and middle scintillation counters. The telescope window that is 
defined by a 2 cm wide finger paddle was centered on the 2.8 cm wide central strip of the RPC. The detailed experimental arrangement of RPCs, 
scintillation counters and electronic circuit are shown in figure~\ref{fig:6}.

\begin{figure}[ht]
        \centering
         \includegraphics[width=1 \textwidth]{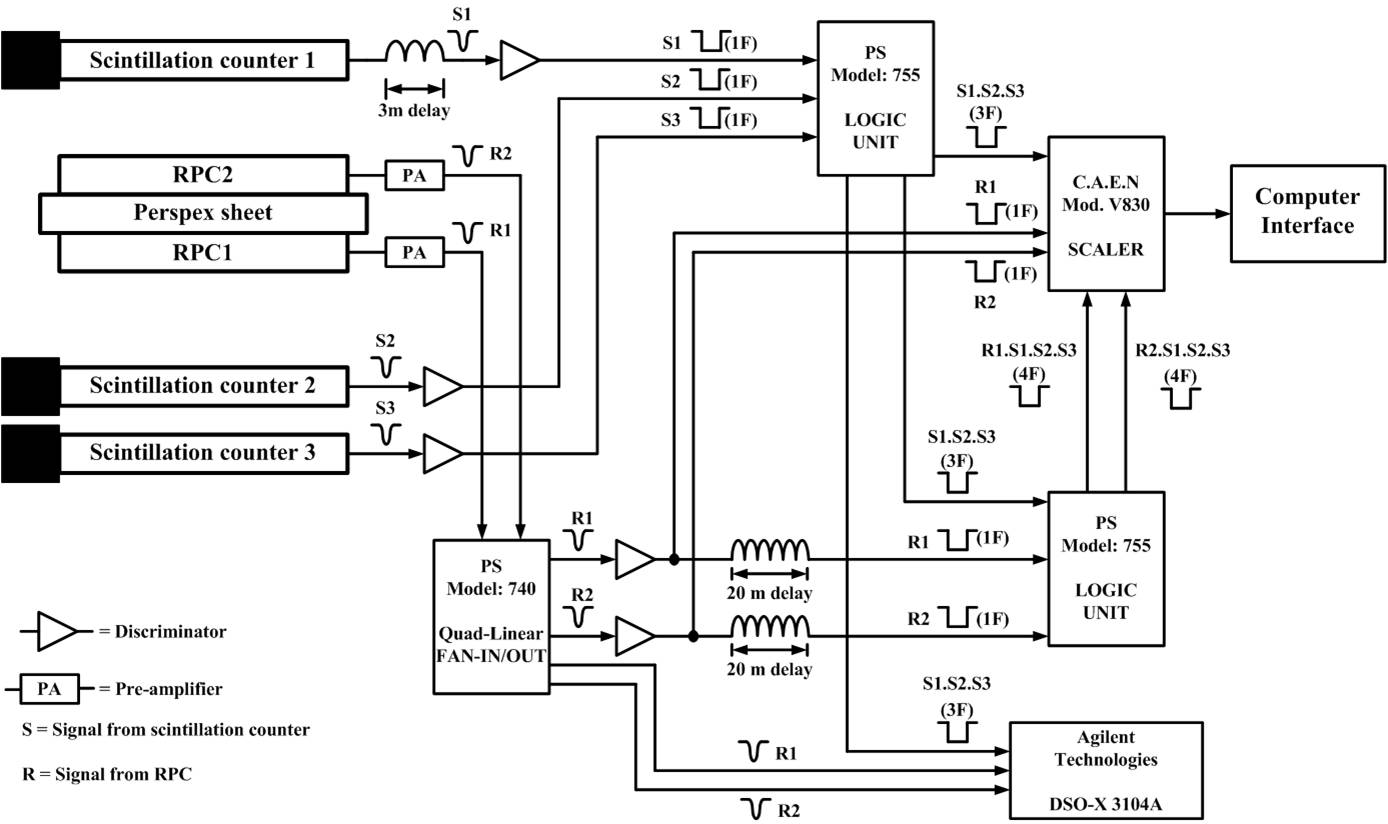}
          \caption{Schematic diagram of the experimental setup developed to test the RPCs.}
\label{fig:6}
\end{figure}
}

\subsection{Test results}
\label{Sec:Test results}

\vskip 0.2cm

{
\noindent {\bf Voltage-current characteristics:}

\vskip 0.1cm

\noindent
The simple electrical equivalent circuit representation of a single gap RPC is shown in figure~\ref{fig:7}. The RPC gas gap is represented
by a parallel combination of spacer resistance and gas ionization volume of the gap (represented by a Zener diode)~\cite{Satyanarayana}. 
Therefore for a given bulk resistivity of the spacers, the current flowing in the circuit depends on the bulk resistivity of the RPC 
electrodes.

\begin{figure}[ht]
        \centering
        \includegraphics[width=0.35\textwidth]{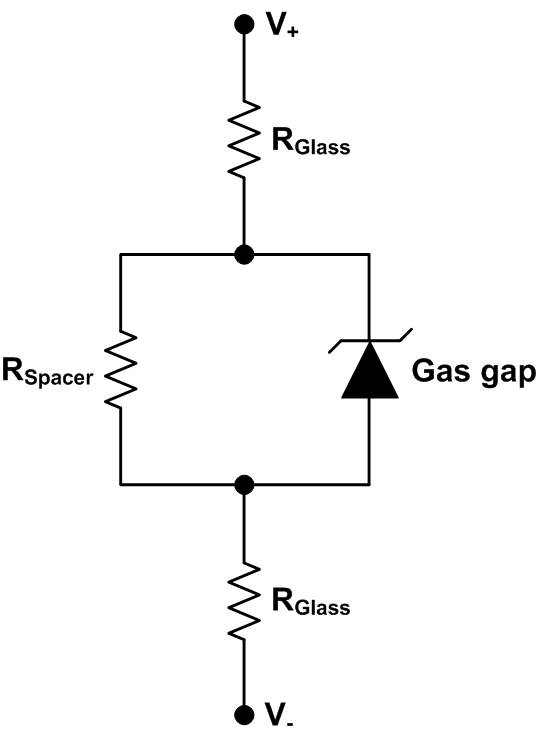}
        \vskip -0.2cm
        \caption {Electrical equivalent circuit representation of a single gap RPC.}
\label{fig:7}
\end{figure}

The voltage-current characteristics of the RPCs were measured using C.A.E.N Mod. N471A, 2 channel HV Power Supply. The current resolution 
of the module is 1 nA. The currents drawn by the RPCs as a function of applied voltage are shown in figure~\ref{fig:8}. A-RPCs were found to 
draw lower bias currents compared to S-RPCs.  This could be because of Asahi glass showed larger bulk resistivity compared to Saint-Gobain
glass.

\begin{figure}[ht]
        \centering
         \includegraphics[width=0.55 \textwidth]{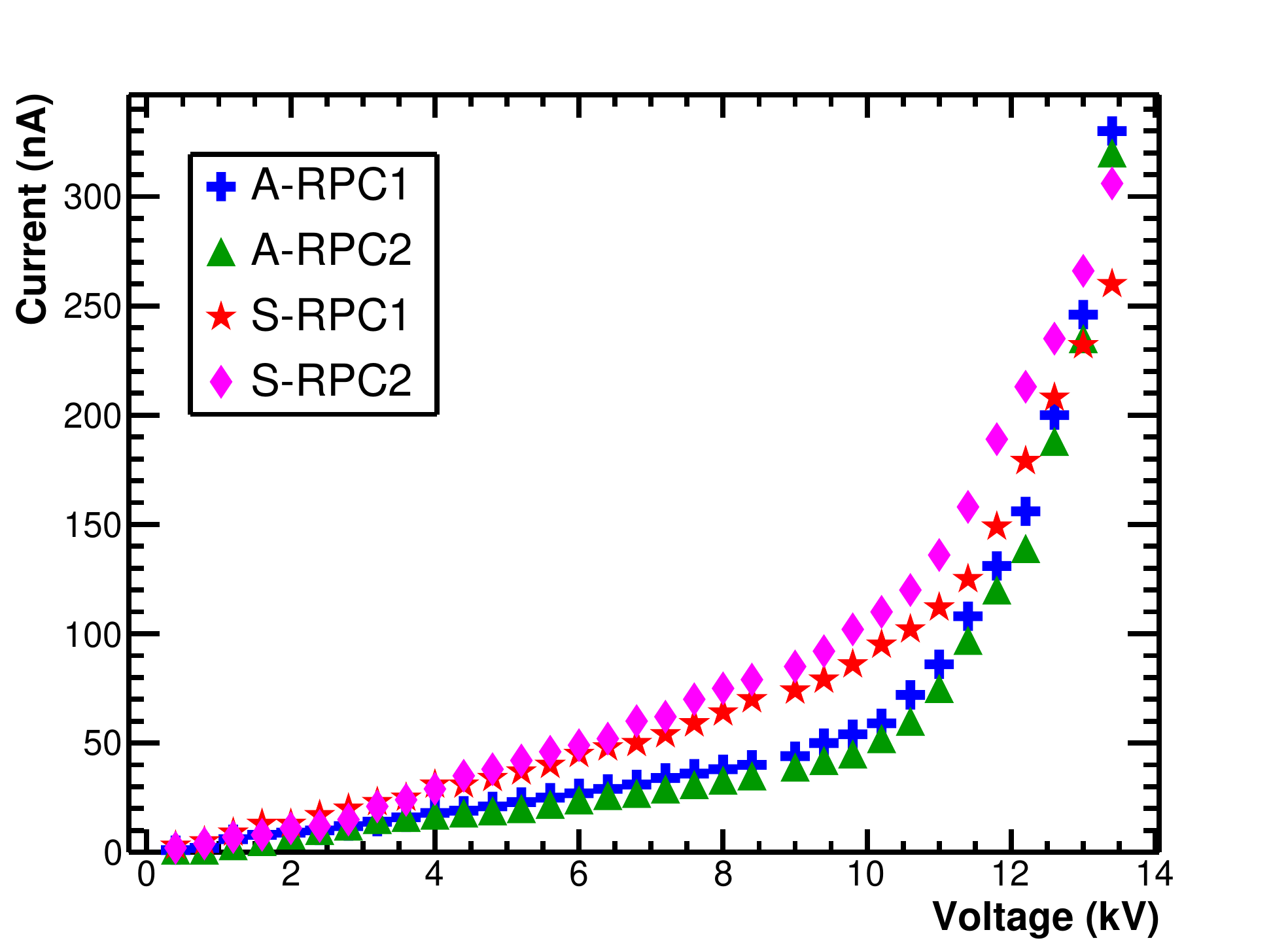}
         \vskip -0.3cm
          \caption{Currents drawn by the RPCs as a function of voltage.}
\label{fig:8}
\end{figure}

\vskip 0.3cm

\noindent {\bf Efficiency studies:}

\vskip 0.1cm

\noindent
The normal vector component of the electric field displacement is continuous at the electrode/gas interface of the RPC. This boundary condition 
is expressed as:
\begin{equation}
{\bf D} = \epsilon_{p} {\bf {E}_{p}} = \epsilon_{g} {\bf {E}_{g}},
\label{eq:3.1} 
\end{equation}

\noindent
where, $\epsilon_{p}$ and $\epsilon_{g}$ are the permittivities, and ${\bf {E}_{p}}$ and ${\bf {E}_{g}}$ are the electric fields of the 
electrode plate and the gas gap, respectively~\cite{Griffiths, Morales_et_al}. From equation~\ref{eq:3.1}, it indicates that the RPCs made out 
of electrodes with larger relative permittivity can be operated at lower bias voltages.

The efficiencies of RPCs were measured using the cosmic-ray muons. The experimental arrangement and electronic circuit for measuring the 
efficiencies of RPCs is shown in figure~\ref{fig:6}. The RPC analog pulses were amplified with indigenously developed pre-amplifier for the 
INO-ICAL RPCs. The gain of the pre-amplifier is $\sim$70. The applied discriminator threshold for the amplified RPC pulses was --20 mV. The 
efficiency measurements on the RPCs were performed using C.A.E.N Mod. V830 Scaler. The telescope trigger pulse was recorded as 3-fold pulse. 
The efficiency is defined as the ratio of the number of coincident pulses of the RPC strip with that telescope trigger (i.e., 4-fold pulses) 
to the number of trigger pulses (3-fold pulses). This definition is written as:

\begin{equation}
 {\text{Efficiency}} = {\frac{4\text{-}{\text{fold}} \ (4\text{F}) \ \text{rate}}{3\text{-}\text{fold} \ (3\text{F}) \ \text{rate}}} \times 100 \%,
\label{eq:3.2}
\end{equation}

\noindent
where, $3\text{F}$ is the telescope trigger pulse, which is generated by time coincidence of three scintillation counters, and $4\text{F}$ is 
the coincidence pulse of the RPC under test and the telescope trigger pulse. 

The measured efficiencies of the RPCs as a function of applied high voltage are shown in figure~\ref{fig:9}. The four RPCs showed greater than 
95\% efficiencies on the plateau. It is observed that the knee of efficiency plateau of A-RPCs starts at 10.8 kV, whereas that of S-RPCs 
starts at 12.0 kV. Therefore, A-RPCs can operate at 1.2 kV lower bias voltage in comparison to S-RPCs. These results are in consistent with 
equation~\ref{eq:3.1}.

\begin{figure}[ht]
        \centering
         \includegraphics[width=0.55 \textwidth]{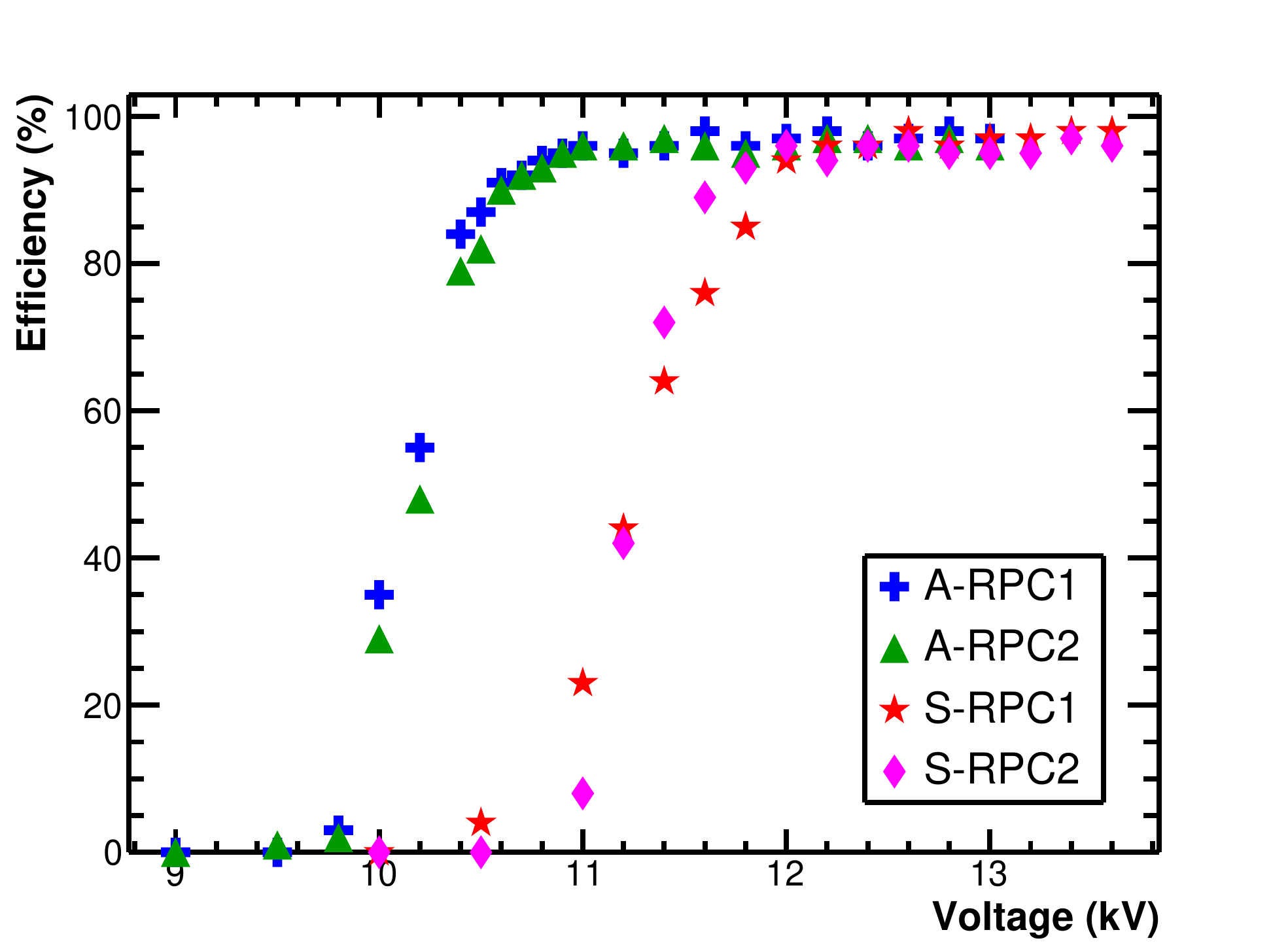}
         \vskip -0.3cm
          \caption{Efficiencies of the RPCs as a function of high voltage.}
\label{fig:9}
\end{figure}

\vskip 0.5cm

\noindent {\bf Time resolutions of the RPCs:}

\vskip 0.1cm

\noindent
From equation~\ref{eq:3.1}, it is clear that for a given applied voltage, the RPC made out of electrodes with larger relative permittivity 
will generate a larger electric field in the gas gap. Therefore, the drift velocity of avalanche electrons will be greater, which leads to 
better time resolutions~\cite{Riegler_Lippmann, Tapasi_Subhasis}.

The RPCs were operated at 12.0 kV, where they showed greater than 95\% efficiency, and measured their time resolutions as followed. 
As shown in figure~\ref{fig:6}, the pulse from scintillation counter 1 (S1) is delayed for about 15 ns (3 m), using RG174 coaxial cable, and 
is used to produce the trigger pulse. Using this trigger pulse, the RPC pulses were recorded on the Agilent DSO-X 3104A. Therefore, here, S1 
is the start counter (T0) to detect the RPC pulses. The time differences between the trigger pulse and RPC pulse were noted on around 1100 
triggers and plotted the time distribution histogram. The plotted histograms are shown in figure~\ref{fig:10}. The obtained sigma from the 
histogram is the quadratic sum of resolution of RPC and resolution of S1, i.e.,

\begin{equation}
 \sigma = \sqrt{\sigma^{2}_{RPC} + \sigma^{2}_{S1}} \implies \sigma_{RPC} = \sqrt{\sigma^{2} - \sigma^{2}_{S1}},
\label{eq:3.3}
\end{equation}

\noindent
where, $\sigma$ is the resolution obtained from the histogram, $\sigma_{RPC}$ is the resolution of RPC and $\sigma_{S1}$ is the resolution
of S1 = 0.74 ns. 

Using equation~\ref{eq:3.3}, the time resolutions of A-RPC1, A-RPC2 are 1.64, 1.53 ns, and that of S-RPC1, S-RPC2 are 2.21, 2.17 ns, 
respectively. The time resolutions of A-RPC1 and S-RPC1 at their operating voltages 11 kV and 12.2 kV are 1.71 and 2.14 ns, respectively. 
The measured pre-amplifier jitter was a few tens of ps and that is much lower than the time resolution of RPC.

\begin{figure}[ht]
        \centering
         \includegraphics[width=1 \textwidth]{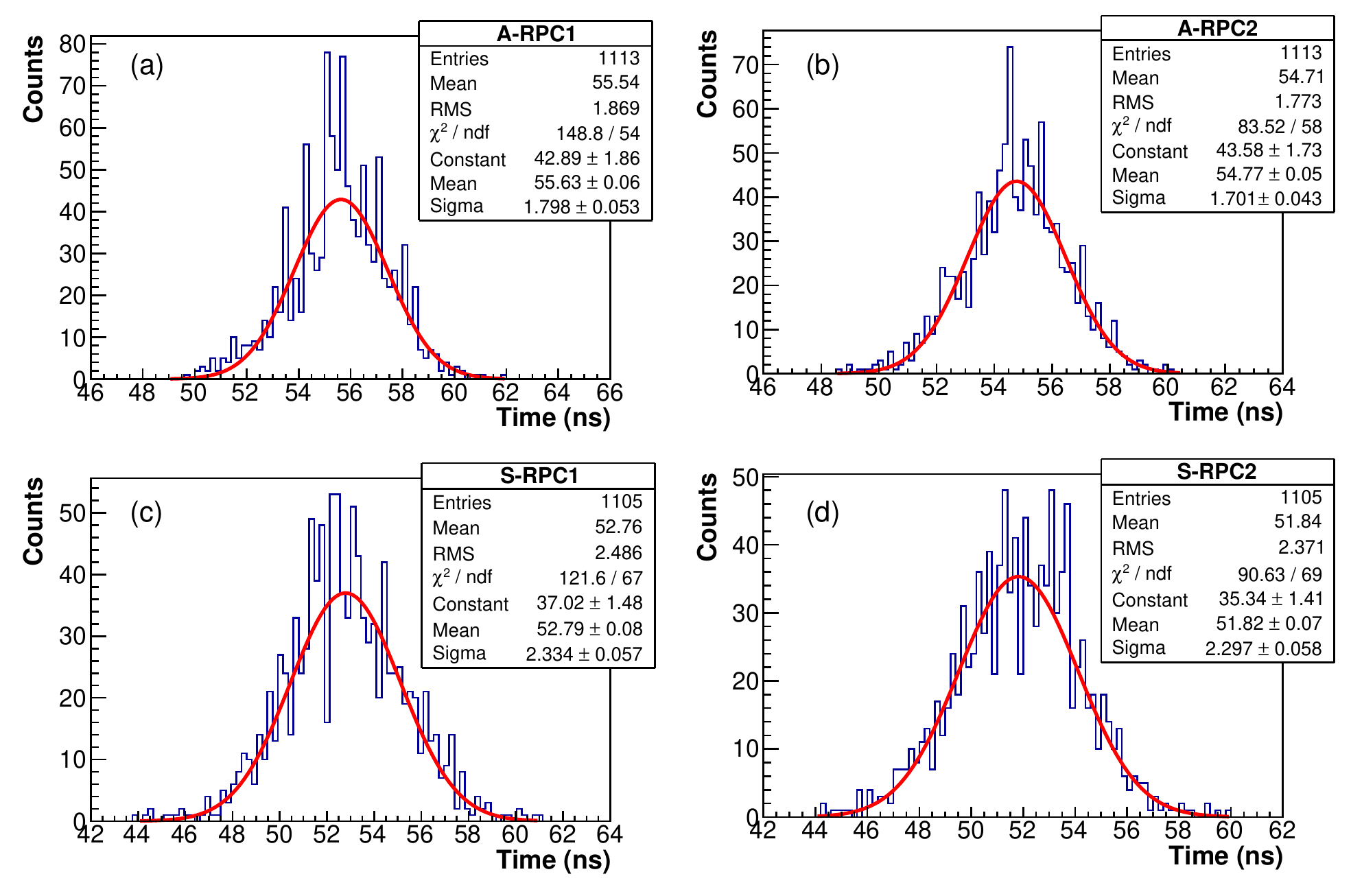}
         \vskip -0.3cm
          \caption{Time resolutions of the RPCs at 12 kV: (a), (b) A-RPCs; (c), (d) S-RPCs.}
\label{fig:10}
\end{figure}
}
}

\section{Conclusions}
\label{Sec:Conclusions}
{
The bulk resistivity and relative permittivities of Asahi and Saint-Gobain glasses were measured. The measurements showed that bulk resistivity 
and relative permittivity of Asahi glass is larger compared to Saint-Gobain glass. The larger permittivy of Asahi glass could be due to the 
presence of larger sodium (Na) component compared to Saint-Gobain glass. RPC detectors were built using these glasses and their performances 
were studied. Asahi RPCs were found to draw lower bias currents compared to Saint-Gobain RPCs. The knee of efficiency plateaus of Asahi RPCs and 
Saint-Gobain RPCs started at 10.8 kV and 12.0 kV, respectively. At a given high voltage, Asahi RPCs showed better time resolutions compared to 
Saint-Gobain RPCs. Our study shows that the RPC made using Asahi glass will be better suited for the INO-ICAL experiment. 
}      

\section*{Acknowledgements}
\label{Sec:Acknowledgements}
{This work was supported by the Department of Atomic Energy (DAE), and the Department of Science and Technology (DST), Government of India. 
The authors would like to gratefully acknowledge the help of their colleagues S. Chavan, V. Asgolkar, Mandar Saraf, R. R. Shinde, V. M. Datar, N. K. Mondal 
and Sanjay Upadhyay at TIFR, Mumbai, and V. Janarthanam, CH. S. R. V. S. Raju, Sudhakar Rao Hari, Gokul Raj R, Soumya Dutta and Logesh 
Karunakaran at IIT Madras.}

\end{document}